\documentclass{rnaastex}
\usepackage{amsmath,color,textcomp,url,graphicx,subfigure}
\usepackage{morefloats}

\newcommand{\kms}{\hbox{km\,s$^{-1}$}}

\newcommand{\gaia}{\emph{Gaia}~DR2}
\newcommand{\msol}{$M_{\odot}$}

\begin{document}

\title{THE LOW-MASS MEMBERS OF THE URSA MAJOR ASSOCIATION}

\author[0000-0002-2592-9612]{Jonathan Gagn\'e}
\affiliation{Plan\'etarium Rio Tinto Alcan, Espace pour la Vie, 4801 av. Pierre-de Coubertin, Montr\'eal, Qu\'ebec, Canada}
\email{jonathan.gagne@montreal.ca}
\author[0000-0001-6251-0573]{Jacqueline K. Faherty}
\affiliation{Department of Astrophysics, American Museum of Natural History, Central Park West at 79th St., New York, NY 10024, USA}
\author{Mark Popinchalk}
\affiliation{Department of Astrophysics, American Museum of Natural History, Central Park West at 79th St., New York, NY 10024, USA}
\affiliation{The Graduate Center, City University of New York, New York, NY 10016, USA}


\section{}

The Ursa~Major association \citep{1992AJ....104.1493E} is a nearby ($\sim$25\,pc) collection of ten F2 to A5-type stars \citep{2003AJ....125.1980K,2018ApJ...856...23G}, aged $414 \pm 23$\,Myr \citep{2015ApJ...813...58J}, distributed in a small core mostly consisting of stars in the Big Dipper asterism, with a scatter in Galactic positions ($XYZ$) of about 1--3\,pc. It is accompanied by a stream of 42 known stars \citep{2003AJ....125.1980K} with a wider range in spectral types and a much wider $XYZ$ scatter of 25--40\,pc. The core is slightly better localized in Galactic velocities $UVW$ with a scatter of 1--3\,\kms, compared with 2.5--4\,\kms\ for the stream members. Despite it being the nearest open cluster to the Sun, its population of low-mass members is still largely missing (only 3 K-type stars and 1 M-type star are part of the stream), likely because they have been scattered away over time. Even most recent surveys based on URAT-South \citep{2018AJ....155..176F}, RECONS \citep{2015AJ....149....5W} and \gaia\ \citep{GaiaCollaboration:2018io} only identified two additional M-type candidate members (2MASS~J13445785+5528219 and 2MASS~J12153937+5239088; \citealt{2018RNAAS...2....9G}). Preliminary surveys using only the kinematic portion of the BANYAN~$\Sigma$ algorithm \citep{2018ApJ...856...23G} to identify lower-mass candidates with a larger $XYZ$ spread, but centered on the same Gaussian model of known members in $UVW$, did not allow us to uncover a significant number of low-mass members, leading us to hypothesize they might have been scattered beyond recovery.

A recent study by \cite{2019AJ....158..122K} used a hierarchical clustering algorithm applied to \gaia\ data to uncover over-densities in kinematic observable space (sky position, proper motion, and parallax) within 1\,kpc and 30\textdegree\ of the Galactic plane. They uncovered 1640 localized or extended over-densities, the latter of which they called strings, and estimated isochronal ages for all of them. Because their method is based in direct observables rather than $XYZUVW$ --the latter three cannot be calculated for the majority of \gaia\ entries which lack a radial velocity measurement-- they did not identify over-densities in the immediate Solar neighborhood, within about 70\,pc around the Sun, except for the very dense Hyades. For this reason, they have not recovered any of the classical members of Ursa~Major, neither the core or stream, as they all lie within this neighborhood.

\begin{figure*}
  \centering
  \subfigure[$XY$]{\includegraphics[width=0.5\textwidth]{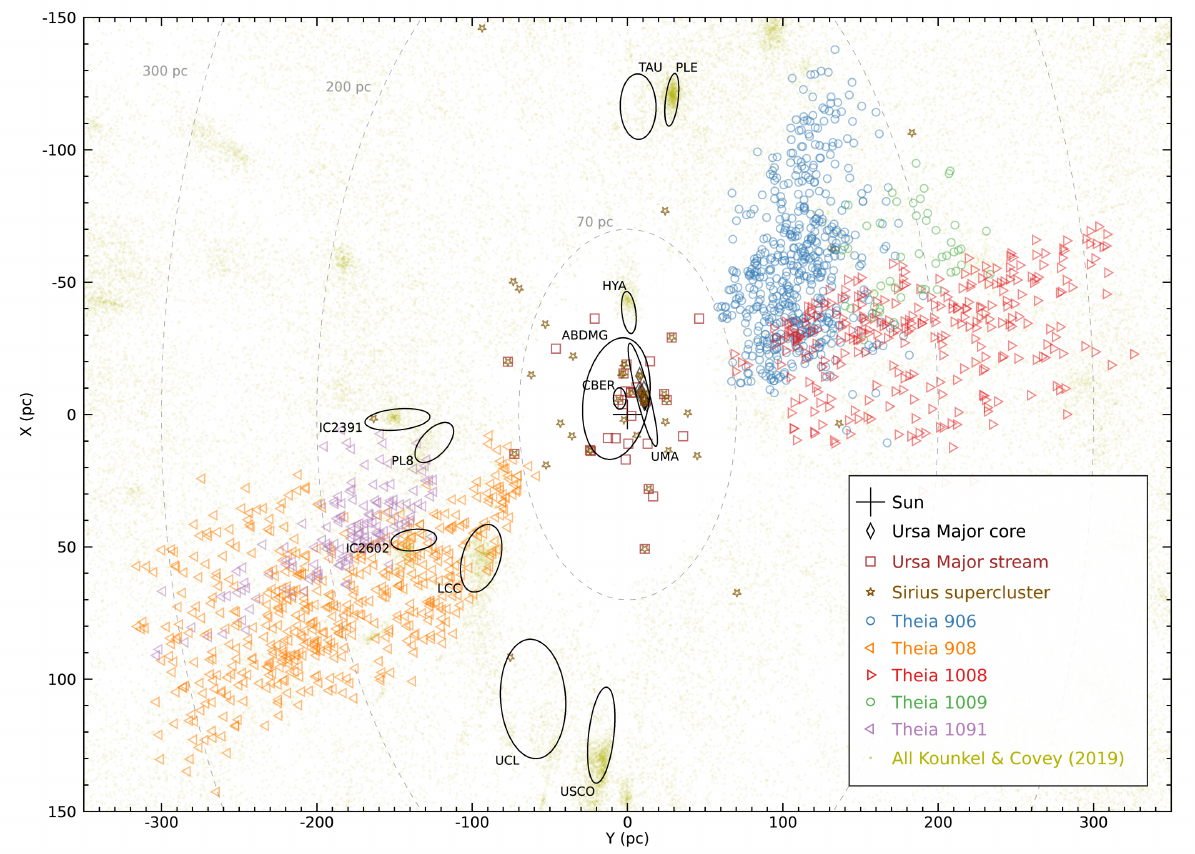}\label{fig:XY}}
  \subfigure[$UV$]{\includegraphics[width=0.5\textwidth]{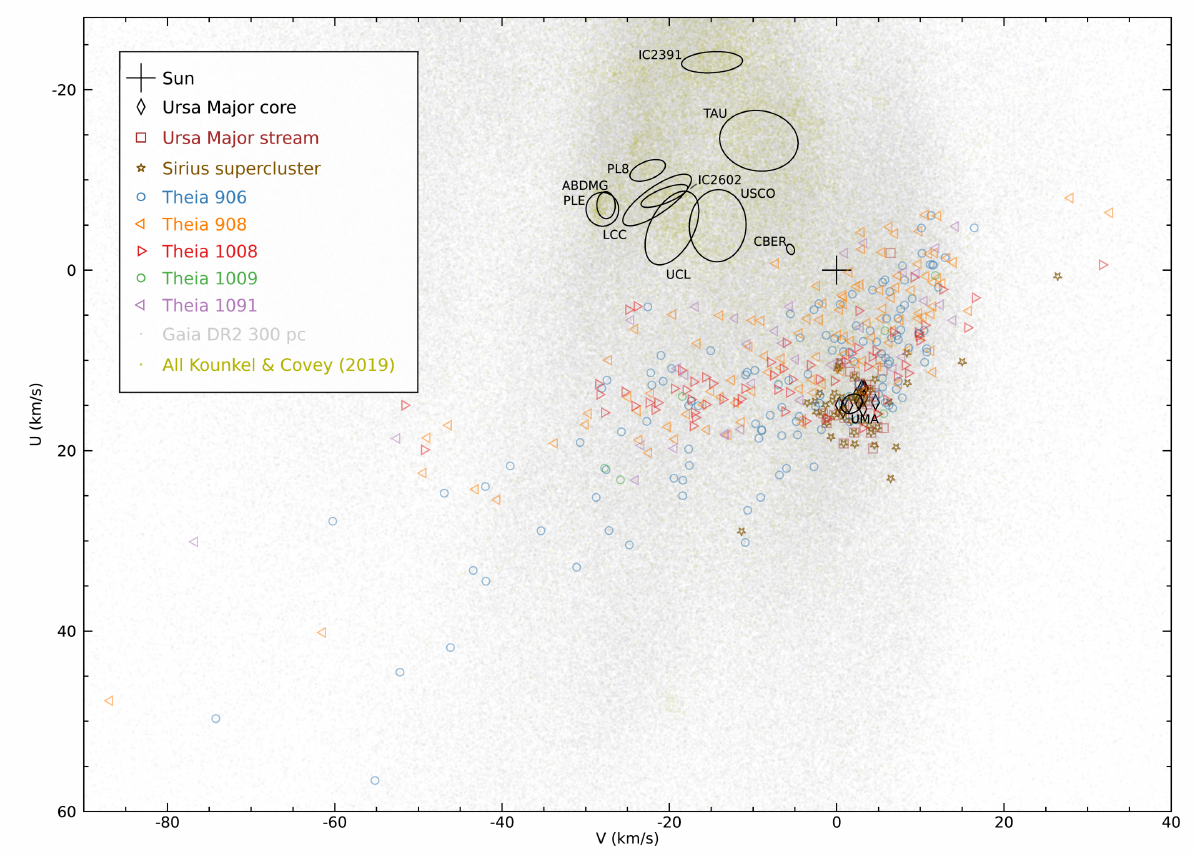}\label{fig:UV}}
  \caption{Galactic positions $XY$ and space velocities $UV$ of the \cite{2019AJ....158..122K} groups that may constitute tidal tails to the Ursa~Major group. They are located between $\sim$70\,pc and $\sim$330\,pc, and have space velocities similar to Ursa Major although more scattered, especially in the direction of Galactic rotation ($V$). Ellipses represent the multivariate Gaussian models of BANYAN~$\Sigma$, and indicate the loci of members for IC~2391, Platais~8, IC~2602, Lower Centaurus Crux (LCC), Upper Centaurus Lupus (UCL), Upper Scorpius (USCO), the Taurus-Auriga star-forming region (TAU), the Pleiades association (PLE), the Hyades association (HYA), the AB~Doradus moving group (ABDMG), the Ursa~Major association (UMA), and the Coma~Berenices association (CBER). For more details on these models and associations, see \cite{2018ApJ...856...23G}. Data behind figures is available as online-only material.}
  \label{fig:kinematics}
\end{figure*}

While inspecting the nearest ($<$\,200\,pc) and youngest ($<$\,1\,Gyr) 73 groups identified by \cite{2019AJ....158..122K}, we noticed that five of them (Theia~906, 908, 1008, 1009 and 1091), totaling 1599 stars, clustered around a similar $UVW$ distribution to the known Ursa~Major core and stream, while extending spatially along two large tails that reach out to almost 350\,pc from the Sun (see Figure~\ref{fig:kinematics}). All but Theia~1091 are designated as strings. Despite their much wider distribution in space velocities (18, 33, and 11\,\kms\ in $U$, $V$ and $W$ respectively), they have been assigned isochronal ages in the range 503--849\,Myr by \cite{2019AJ....158..122K}, all of which could be consistent with real members of Ursa~Major given that these isochronal ages are based on models with significant systematics (e.g., see the discussion of \citealt{2018ApJ...861L..13G}), and have not been corrected for interstellar extinction. The wider spread in space velocities, especially along the direction of Galactic rotation $V$, could indicate that these arms are tidal tails well into the process of dissipation, but not exactly beyond recovery as we had previously hypothesized. It is interesting to note that the proposed Sirius supercluster also identified by \cite{1992AJ....104.1493E} displays a similar distribution in $UVW$ space to the Theia groups from \cite{2019AJ....158..122K} discussed here, and could be related although it may be a contaminated sample--\cite{1992AJ....104.1493E} noted that it seems to consist of some stars aged $\sim$\,630\,Myr and others $\sim$\,1\,Gyr.

The \gaia\ absolute $G$ magnitude versus $G - G_{\rm RP}$ color diagram of the Theia groups discussed here seems consistent with a coeval, $\sim$500\,Myr-old stellar population: it ends abruptly at the upper main sequence around $G - G_{\rm RP} = 0$, and its most massive component, HD~199713, is a B9 star \citep{2003AJ....125..359W}, corresponding to an estimated mass of $\sim$2.8\,\msol\ from the spectral type--mass relations of \cite{2013ApJS..208....9P}. Other mass estimates for HD~199713 based on a comparison of its \gaia\ photometry to models or empirical relations from eclipsing binaries range from $2.9_{-0.4}^{+0.3}$\,\msol\ \citep{2019AA...628A..94A} to $3.0 \pm 0.4$\,\msol\ \citep{2019AJ....158..138S}. These mass estimates correspond to stars that spend 460--580\,Myr on the main sequence according to the MIST models \citep{2016ApJ...823..102C}, consistent with current age estimates for Ursa~Major. The lowest-mass portion of the color-magnitude diagram contains 990 stars with $G - G_{\rm RP}$ between 1.0 and 1.4\,mag, corresponding to spectral types M0--M5, and are not significantly over-luminous compared to the distribution of older field stars, also consistent with the Ursa~Major age. Both the Ursa~Major stream defined by \cite{2003AJ....125.1980K} and the Sirius supercluster described by \cite{1992AJ....104.1493E} have similar color-magnitude diagram sequences with consistent main-sequence turn-off points. Only one Sirius supercluster star is bluer, $\beta$~Aur (spectral type A1, see \citealt{2003AJ....126.2048G}), but this is likely due to it being an eclipsing binary. The mass of its individual components have been measured by \cite{1995BICDS..47....9B} at $2.35 \pm 0.03$\,\msol\ and $2.27 \pm 0.03$\,\msol, and are therefore less massive than HD~199713 and consistent with the quoted age of Ursa~Major.

The aforementioned \cite{2019AJ....158..122K} groups contain 20 white dwarfs based on their position in the \gaia\ color-magnitude diagram: 16 of them are clearly too old for bearing any relation to Ursa~Major based on a comparison with the Montr\'eal C/O core cooling tracks\footnote{Computed similarly as the pure C core models described in \cite{2001PASP..113..409F}, and available at \url{http://www.astro.umontreal.ca/~bergeron/CoolingModels/}}. Further, all of these contaminant white dwarfs are located within 200\,pc, suggesting that the more distant ones have simply not been uncovered because of their faintness. Combining this number with local number space densities for white dwarfs ($4.49 \pm 0.38 \times 10^{-3}$ objects\,pc$^{-3}$; \citealt{2018MNRAS.480.3942H}) and main-sequence stars ($98.4 \pm 6.8 \times 10^{-3}$ objects\,pc$^{-3}$; \citealt{2012ApJ...753..156K}) suggests we might expect a total of $\sim$520 main-sequence false-positive members, or a contamination rate of about $\sim$33\%. We note that most of the white dwarf interlopers are located in Theia~906 (6 interlopers) and Theia~908 (also 6 interlopers), indicating they may be more contaminated -- consistent with their wider $UVW$ distributions. The remaining are in Theia~1008 (3 interlopers) and Theia~1091 (1 interloper). Four additional white dwarfs (WDJ235833.30+510815.54, WDJ233848.60+404803.82, WDJ232227.23+394456.36, and WDJ121521.73$-$523646.04) have temperatures in the range 8000\,K--24000\,K, masses in the range 1.00--1.27\,\msol\ \citep{2019MNRAS.482.4570G}, and total (main-sequence plus cooling) ages in the range 400--800\,Myr, based on non-dereddened \gaia\ color-magnitude positions.

Most of the members in these candidate tidal tails are distributed between $-$20 to +20\textdegree\ in Galactic longitude, indicating that there might not be many candidates left to be recovered beyond the $-$30 to +30\textdegree\ range that was surveyed by \cite{2019AJ....158..122K}. Stars at the upper main sequence of the color-magnitude diagram have a range of distances from 73\,pc to 330\,pc, compared to 67--330\,pc for the full sample: this indicates no obvious distance-dependent contamination in their sample of lower-mass stars. It can however be expected that more members may be left to uncover within 70\,pc of the Sun, where the \cite{2019AJ....158..122K} algorithm was less sensitive. We have observed no clear correlation between the spatial $XYZ$ and kinematic $UVW$ distribution of members, meaning that any search for nearby members will likely require using a very wide net in $UVW$ space, and may suffer a high rate of contamination despite the unusual kinematics of Ursa~Major compared with the Solar neighborhood. This also explains why our previous searches based on the $UVW$ distribution of the core members failed to uncover anything.

More work is clearly needed to disentangle the relations between the different groups discussed here, and to assess whether the large structures recovered by \cite{2019AJ....158..122K} are really related to Ursa~Major, and where the \cite{1992AJ....104.1493E} Sirius supercluster, and the \cite{2003AJ....125.1980K} Ursa~Major stream, fit into that picture. It is likely, for example, that some members at the extremes of the $UVW$ distributions of Theia~906 and 908 suffer from low-quality measurements or correspond to contaminants. Using metrics of stellar activity and rotation periods from the {\it TESS} mission \citep{2015JATIS...1a4003R} will be useful to corroborate membership although only for the nearest members (e.g., see \citealt{2019AJ....158...77C}), and a more detailed study of individual members will be useful to further investigate the relation between the \cite{2019AJ....158..122K} groups discussed here and Ursa~Major. \emph{Gaia}~DR3 will be useful in this investigation as it will likely contain radial velocity measurements for several stars discussed here. It will also be interesting to determine why Ursa~Major might have much larger tidal tails with a wider space velocity spread than the similarly-aged Coma~Ber \citep{2014AA...566A.132S} which tidal tails were recently discovered by \cite{2019ApJ...877...12T}. Studying the four white dwarfs discussed above in more details will also be useful to better age-date this population.

\acknowledgments

We would like to thank Eric E. Mamajek and Andrew W. Mann for useful comments and discussion. This research made use of data from the ESA mission \emph{Gaia}, processed by the {\it Gaia} Data Processing and Analysis Consortium whose funding has been provided by national institutions, in particular the institutions participating in the {\it Gaia} Multilateral Agreement.

\software{BANYAN~$\Sigma$ \citep{2018ApJ...856...23G}.}

\newpage

\end{document}